\documentclass[%
 reprint,
 amsmath,amssymb,
 aps,
]{revtex4-2}

\usepackage{graphicx}
\usepackage{dcolumn}
\usepackage{bm}
\usepackage[mathlines]{lineno}
\usepackage{xcolor}

\begin{document}

\title{Orbital Hanle magnetoresistance in a 3$d$ transition metal}

\author{Giacomo Sala}
\affiliation{Department of Materials, ETH Zurich, Hönggerbergring 64, 8093 Zurich, Switzerland}
\author{Hanchen Wang}
 \affiliation{Department of Materials, ETH Zurich, Hönggerbergring 64, 8093 Zurich, Switzerland}
\author{William Legrand}
 \affiliation{Department of Materials, ETH Zurich, Hönggerbergring 64, 8093 Zurich, Switzerland}
 \author{Pietro Gambardella}
 \affiliation{Department of Materials, ETH Zurich, Hönggerbergring 64, 8093 Zurich, Switzerland}

\begin{abstract}

The Hanle magnetoresistance is a telltale signature of spin precession in nonmagnetic conductors, in which strong spin-orbit coupling generates edge spin accumulation via the spin Hall effect. Here, we report the existence of a large Hanle magnetoresistance in single layers of Mn with weak spin-orbit coupling, which we attribute to the orbital Hall effect. The simultaneous observation of a sizable Hanle magnetoresistance and vanishing small spin Hall magnetoresistance in BiYIG/Mn bilayers corroborates the orbital origin of both effects. We estimate an orbital Hall angle of 0.016, an orbital relaxation time of 2~ps and diffusion length of the order of 2~nm in disordered Mn. Our findings indicate that current-induced orbital moments are responsible for magnetoresistance effects comparable to or even larger than those determined by spin moments, and provide a tool to investigate nonequilibrium orbital transport phenomena.
\end{abstract}

\maketitle

\textit{The original manuscript has been published with DOI 10.1103/PhysRevLett.131.156703 by the American Physical Society, which ows the copyright of the published article. Publication on a free-access e-print server of files prepared and formatted by the authors is allowed by the American Physical Society. The authors own the copyright of these files.}

\vspace{0.5cm}

The angular momentum induced by electric currents in thin films is at the origin of several types of magnetoresistance. In ferromagnet/nonmagnet bilayers, the spin current generated by the spin Hall effect in the nonmagnetic conductor interacts with the magnetization in the ferromagnet, giving rise to the spin Hall magnetoresistance (SMR) \cite{Nakayama2013,Chen2013,Vlietstra2013,Althammer2013,Isasa2014,Avci2015,Avci2015a,Kim2016a,Zhang2019a}. In the SMR scenario, the bilayer resistance is maximum (minimum) when the magnetization $\mathbf{M}$ is perpendicular (parallel) to the polarization $\boldsymbol{\zeta}$ of the spin current. When $\mathbf{M} \parallel \boldsymbol{\zeta}$, the spin current is reflected at the interface and is converted into a transverse charge current by the inverse spin Hall effect. This conversion yields a lower resistance. In contrast, when $\mathbf{M} \perp \boldsymbol{\zeta}$, the spin current is absorbed, the spin-to-charge conversion is inhibited, and the resistance is maximum. A similar magnetoresistance occurs when the spin polarization is generated by the Rashba-Edelstein effect \cite{Zhang2015d,Grigoryan2014,Nakayama2016}. 

Similar to the SMR, the Hanle magnetoresistance (HMR) also originates from the modulation of the spin-charge interconversion by the inverse spin Hall effect \cite{Johnson1985,Dyakonov2007,Velez2016,Wu2016,Li2022a}. However, contrary to the SMR, the HMR appears in single nonmagnetic layers when the interfacial exchange field due to the magnetization is replaced by an external magnetic field $\mathbf{B}$. As sketched in Fig. 1, the spins accumulated at the edges of the nonmagnetic layer by the spin Hall effect precess about $\mathbf{B}$, and the net spin polarization decreases because of the combination of the precession and electron diffusion. The consequent attenuation of the inverse spin Hall effect results in a higher resistance. Since the Larmor precession and spin dephasing are the core ingredients of the Hanle effect, the HMR is large when $\omega \tau > 1$, with $\omega$ and $\tau$ the Larmor frequency and the spin relaxation time, respectively. If $\tau \approx$ 1 ps, magnetic fields of the order of a few Tesla are required to observe the HMR. Overall, the SMR and HMR provide a powerful means to understand the generation and transport of angular momentum in thin films and across interfaces.

Both magnetoresistances scale as the square of the spin Hall angle $\theta^2$, which quantifies the strength of the charge-to-spin interconversion. Therefore, heavy elements with large spin-orbit coupling such as Pt and Ta are known to generate strong SMR and HMR. However, recent theories \cite{Go2017,Go2018a,Salemi2019,Bhowal2020,Johansson2020,Salemi2022b} and experiments \cite{Choi2021a} have shown that electric currents can induce a large orbital accumulation in both light and heavy metals as a consequence of the orbital Hall and orbital Rashba-Edelstein effects. The orbital angular momentum can source spin-orbit torques \cite{Ding2020,Kim2021a,Lee2021b,Lee2021c,Sala2022a,Dutta2022,Hayashi2023}
and contribute to the magnetoresistance \cite{Ko2020,Ding2021,Ding2022}. However, whether the orbital momentum can induce the orbital analog of the SMR and HMR is an open question \cite{Ding2021}. More generally, the physics underlying the bulk and interfacial transport of orbital momentum, the interaction of orbital moments with the magnetization and magnetic fields, and the associated time- and lengthscale remain poorly understood.

\begin{figure}
\includegraphics[scale=0.5]{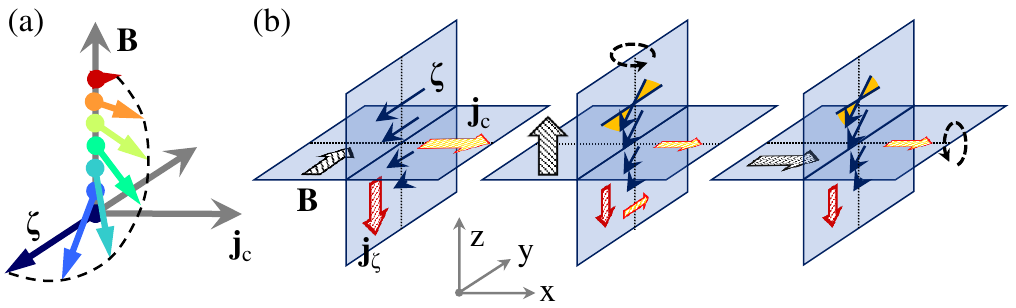}
\caption{(a) Schematic representation of the Hanle effect. The angular momentum $\boldsymbol{\zeta}$ induced by an electric current $\mathbf{j}_{\textrm{c}}$ via the spin or orbital Hall effect precesses about the magnetic field $\mathbf{B}$. The net  $\boldsymbol{\zeta}$ decreases because electrons travelling along different paths accumulate different phases. (b) Hanle effect in three different configurations. Left: when $\mathbf{B} \parallel \boldsymbol{\zeta}$, no precession occurs, and the spin (orbital) current $\mathbf{j}_{\boldsymbol{\zeta}}$ polarized along $-y$ is converted by the inverse spin (orbital) Hall effect into a charge current $\mathbf{j}_{\textrm{c}}$ along $x$. The resistance is thus minimum. Middle and right: $\mathbf{B}$ induces the precession of $\boldsymbol{\zeta}$, reducing  $\zeta_{\textrm{y}}$, and suppressing both $\mathbf{j}_{\boldsymbol{\zeta}}$ and $\mathbf{j}_{\textrm{c}}$. The resistance is maximum. When $\mathbf{B} \parallel \mathbf{z}$ (middle), an additional charge current flows along $y$ because of the field-induced $\zeta_{\textrm{x}}$. This current induces the transverse HMR.}
\label{fig:fig1}
\end{figure}

Here, we demonstrate the existence of a sizable HMR in single layers of Mn. The longitudinal and transverse HMR of Mn can be as high as $6.5\cdot10^{-5}$ and $2\cdot10^{-5}$, respectively, which are comparable to or even larger than the HMR of Pt and Ta \cite{Velez2016,Wu2016,Li2022a}. The magnitude of the HMR and the weak spin-orbit coupling of Mn suggest that this magnetoresistance is driven by orbital moments. The orbital origin of the HMR is further supported by the very small SMR $\approx 6.4\cdot 10^{-6}$ found in BiYIG/Mn bilayers, in which orbital moments do not interact directly with the magnetization of BiYIG \cite{Go2021}. The analysis of the field and thickness dependence of the HMR in Mn provides insight into the orbital relaxation time and diffusion length. Our findings reveal the existence of a sizable magnetoresistance of orbital origin in a light transition-metal element, which provides a new tool to investigate nonequilibrium orbital transport phenomena.

\begin{figure}
\includegraphics[scale=1]{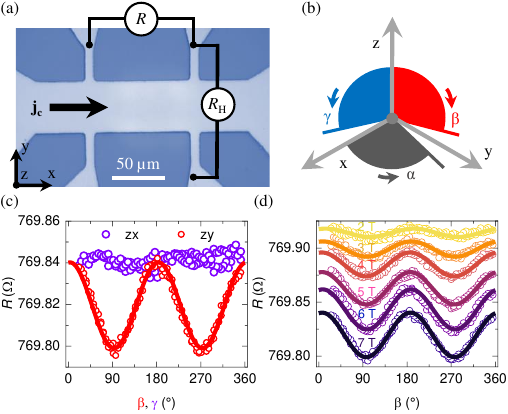}
\caption{(a) Optical image of a Hall bar device and sketch of the measurement configuration. $R$ and $R_{\textrm{H}}$ are the longitudinal and transverse resistance, respectively. (b) Coordinate system showing the angle of the applied field in the $xy$, $zx$, and $zy$ planes.
(c) Longitudinal resistance measured in Mn(6) by $zx$ and $zy$ angle scans in a field of 7~T. (d) Longitudinal resistance in the same device during $zy$ angle scans at increasing magnetic fields. The solid lines in (c, d) are fits to a $\cos^2(\beta)$ function. The angle scans are limited to the $[25^{\circ}-355^{\circ}]$ range because of technical constraints.}
\label{fig:fig2}
\end{figure}

We studied the HMR in Mn($t_{\textrm{Mn}}$) layers with variable thickness $t_{\textrm{Mn}} = 4-40$ nm by performing both angular measurements of the magnetoresistance at constant magnetic field and field sweeps along orthogonal directions [Fig. 2(a,b)]. The Mn layers were grown by magnetron sputtering on Si/SiO$_2$ substrates, capped with SiN(8), and patterned in Hall bars by Ar ion etching \cite{SuppPR}.
\nocite{Miller2022,Kahl2004,Caretta2020b,Boakye1992,Skuja2005,Poindexter1995}
In the following, we  generalize the parameters $\boldsymbol{\zeta}$ and $\theta$ to the orbital polarization and orbital Hall angle, respectively, noting that the spin and orbital degrees of freedom are not entirely separable in the presence of spin-orbit coupling \cite{Sala2022a}. 

Figure 2(c,d) shows the room-temperature longitudinal resistance measured in Mn(6) while rotating the magnetic field in the $zx$ and $zy$ planes. The resistance remains constant when the field rotates in the $zx$ plane, which is perpendicular to the current-induced $\boldsymbol{\zeta}$ ($\parallel y$). In contrast, the resistance shows a $\cos^2\beta$ dependence when the field is rotated in the $zy$ plane with the angles defined in Fig.~2(b). In particular, the resistance is maximum when $\mathbf{B} \perp \boldsymbol{\zeta}$ and minimum when $\mathbf{B} \parallel \boldsymbol{\zeta}$. This angular dependence is typical of the HMR, whose longitudinal and transverse components depend on the magnetic field vector $\bm{b} = \bm{B}/B = [b_x, b_y, b_z] = [\cos\alpha\sin\delta, \sin\alpha\sin\delta, \cos\delta]$, with $\delta = \beta, \gamma$, as

\begin{gather}
R = R_0  + \Delta R(1 - b_y^2), \label{eq:R_long} \\
R_{\textrm{H}} = \Delta R_{\textrm{H}}b_z  +  \Delta R\,b_xb_y, \label{eq:R_trans}
\end{gather}

where $R_0$ is the field-independent longitudinal resistance, and $\Delta R$ and $\Delta R_{\textrm{H}}$ are determined by the diffusion and precession of the spin and orbital moments, as described below. The angular symmetry defined by Eqs. \ref{eq:R_long}-\ref{eq:R_trans} is the same as that of the SMR, however the dependence of the longitudinal resistance on the field strength [Fig. 2(d)] and the absence of magnetic layers rule out the SMR as a cause of the observed effects. The ordinary Lorentz magnetoresistance is also excluded by the flat response measured in the $zx$ angle scans [Fig. 2(c)], the non-parabolic dependence of $\Delta R$ and nonlinear dependence of $\Delta R_{\textrm{H}}$ on the magnetic field (see below, Fig. 3), and the nonmonotonic dependence of $\frac{\Delta R}{R_0}$ on the film thickness (see below, Fig. 4(a)). Because these measurements are performed at room temperature, weak antilocalization effects cannot influence the sample resistance \cite{Velez2016}. Finally, the  magnetoresistance in Fig. 2(c,d) cannot be associated with a magnetically-ordered state of Mn because the Néel temperature of bulk Mn is far below room temperature \cite{Murayama1977,Boakye1996}, and our samples do not show any indication of antiferromagnetism \cite{SuppPR}.

To confirm the Hanle origin of the observed magnetoresistance, we performed field scans along the three coordinate axes, as shown in Fig. 3(a,b). The longitudinal resistance increases monotonically with the magnetic field only if the latter is applied along $x$ or $z$, and remains stable when the field is along $y$. The transverse resistance, instead, vanishes when the field is oriented along $x$ or $y$ and follows a sigmoidal curve when the field is along $z$. The dependence of both the longitudinal and transverse resistances on the direction and amplitude of the magnetic field are characteristic fingerprints of the HMR (Fig. 1 and Eqs. \ref{eq:R_long}-\ref{eq:R_trans}) \cite{Johnson1985,Dyakonov2007,Velez2016,Wu2016,Li2022a}. As a reference, we show in Fig. 3(c,d) similar measurements performed in a Pt(5) device fabricated with the same procedure as for Mn. The two materials show the same response to external magnetic fields, although the parabolic (non-saturating) behavior of the longitudinal (transverse) magnetoresistance of Pt indicates different scattering parameters compared to Mn (see below and Table I in Ref.~\onlinecite{SuppPR}). Importantly, the normalized longitudinal HMR, defined as $\frac{\Delta R}{R_0} = [R(7 \, \textrm{T}) - R(0)]/R(0)$, is $6.5\cdot10^{-5}$ and $9.0\cdot10^{-5}$ in Mn(9) and Pt(5), respectively. Whereas the HMR of Pt is in good agreement with earlier reports \cite{Velez2016,Wu2016,Li2022a}, the large HMR of Mn is unexpected for an element with negligible spin-orbit coupling \cite{Wang2014a}. As we discuss below, this HMR provides evidence of physics beyond the spin Hall effect. 

To obtain a quantitative insight into the HMR of Mn, we performed systematic measurements of the magnetoresistance as a function of $t_{\textrm{Mn}}$, as shown in Fig. 4(a). Here, the normalized transverse HMR is calculated as $\frac{\Delta R_{\textrm{H}}}{R_0} = [R_{\textrm{H}}(7 \, \textrm{T}) - R_{\textrm{H}}(0)]/R(0)$. Both the longitudinal and transverse HMR show a nonmonotonic dependence on the film thickness that is typical of diffusive phenomena occurring on the lengthscale of the diffusion length $\lambda$ \cite{Dyakonov2007,Velez2016,Li2022a}. This dependence is described by

\begin{gather}
\frac{\Delta R}{R_0} = 2\theta^2\left[ \frac{\lambda}{t} \tanh \left(\frac{t}{2\lambda}\right) - \mathfrak{Re}\left\{\frac{\Lambda}{t} \tanh \left(\frac{t}{2\Lambda}\right)\right\} \right], \label{eq:Rxx} \\
\frac{\Delta R_{\textrm{H}}}{R_0} = 2\theta^2 \, \mathfrak{Im}\left\{\frac{\Lambda}{t} \tanh \left(\frac{t}{2\Lambda}\right)\right\}, \label{eq:Rxy}
\end{gather}

\begin{figure}
\includegraphics[scale=1]{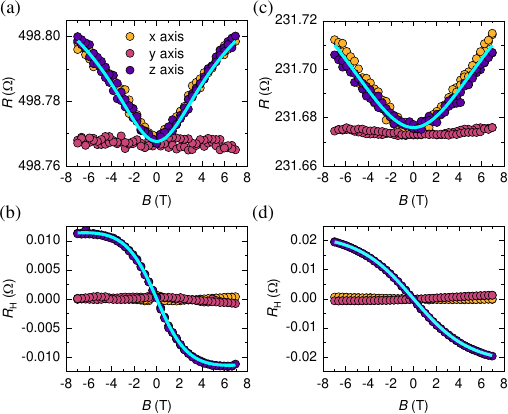}
\caption{(a-b) Longitudinal and transverse resistance of Mn(9) measured during field scans along three orthogonal directions. The contribution from the ordinary Hall effect was subtracted from the transverse resistance measured in the $z$-field-scan. The solid line in (a) is a  fit with shared parameters of Eq. \ref{eq:Rxx} to both the $x$- and $z$-field-scan magnetoresistance. The solid line in (b) is a fit of Eq. \ref{eq:Rxy} to the $z$-field-scan magnetoresistance. (c-d) Same as (a-b) for Pt(5).}
\label{fig:fig3}
\end{figure}

\noindent where $t$ is the film thickness and $\Lambda^{-1} = \sqrt{1/\lambda^2 + i/\lambda_m^2}$. Here, $\lambda_m = \sqrt{D/\omega}=  \sqrt{D\hbar/g\mu_{\textrm{B}}B}$ is an effective length that quantifies the interplay between the electron diffusion and field-induced precession of the angular momentum ($D$, $\hbar$, $g$, and $\mu_{\textrm{B}}$ are the diffusion coefficient, the reduced Planck constant, the Landé g-factor, and the Bohr magneton, respectively). Equations \ref{eq:Rxx}-\ref{eq:Rxy} describe the field and thickness dependence of the HMR. In particular, they predict that in the low-field regime ($\lesssim 2$ T) the longitudinal and transverse resistance increase quadratically and linearly with the magnetic field, respectively, which is consistent with the experimental curves in Fig. 3.

Fitting Eqs.~\ref{eq:Rxx}-\ref{eq:Rxy} to the field dependence in Fig. 3 (both longitudinal and transverse HMR) and the thickness dependence in Fig. 4(a) yields consistent results, as summarized in Table I in Ref.~\onlinecite{SuppPR}. A most striking finding is the similar $\theta$ of Mn and Pt, which are about 0.016 and 0.033, respectively. Whereas $\theta$ of Pt is in line with previous measurements \cite{Velez2016,Tao2018,Manchon2019,Li2022a}, that of Mn is 10 times larger than found in spin-pumping measurements of YIG/Mn bilayers \cite{Du2014,Qu2018}. 
The apparent contradiction between this result and the weak charge-spin conversion efficiency expected for an element with small spin-orbit coupling may be reconciled if we consider the orbital polarization. Theory shows that the orbital Hall effect is much stronger than the spin Hall effect in 3$d$ transition-metal elements \cite{Jo2018,Salemi2022a}.
Because both spin and orbital moments precess in a magnetic field, the orbital momentum should also give rise to a finite HMR. Therefore, the unexpected large HMR in the absence of heavy elements suggests the presence of an important orbital Hall effect.

\begin{figure}
\includegraphics[scale=1]{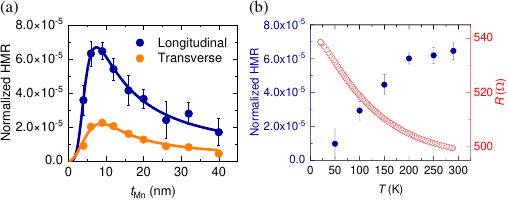}
\caption{(a) Thickness dependence of the normalized longitudinal and transverse HMR. (b) Temperature dependence of the longitudinal HMR (left axis) and longitudinal resistance (right axis) in Mn(9).}
\label{fig:fig4}
\end{figure}

The parameters extracted from the fits give insight into the physics of the orbital transport, which has remained so far elusive. 
We find an orbital conductivity $\sigma_L \approx 55 \, \frac{\hbar}{e}\, (\Omega$cm)$^{-1}$, diffusion length $\lambda \approx$ 2 nm, diffusion coefficient $D \approx$ 2.5$\cdot10^{-6}$ m$^2$/Vs, and orbital relaxation time $\tau = \lambda^2/D \approx$ 2 ps. We note that the estimate of $D$ and $\tau$ is affected by uncertainties on the Landé factor $g$. Whereas $g = 2$ for an electron with spin-only magnetic moment, the Landé factor of an orbital-carrying conduction electron depends on the orbital part of the wavefunction. We thus expect $g \neq 2$, but the exact value remains unknown. The orbital diffusion length that we estimate is five times shorter than the spin diffusion length determined from spin pumping measurements in YIG/Mn \cite{Du2014}. Remarkably, $\lambda$ is significantly smaller than the orbital diffusion length estimated from spin-orbit torque measurements in other 3$d$ metals \cite{Lee2021b,Sala2022a,Hayashi2023}. The estimated orbital Hall conductivity is also significantly smaller than the value predicted by theory for bcc Mn \cite{Jo2018,Salemi2022a}, which is about 9000 $\frac{\hbar}{e}$ $(\Omega$cm)$^{-1}$. In the absence of a clear theoretical understanding of the mechanisms responsible for the orbital quenching and relaxation, we hypothesize a relation between the small orbital Hall conductivity, the short diffusion length, and the disordered crystal structure of sputtered Mn, as discussed below. 

Mn has the largest and most complex unit cell of all transition-metal elements \cite{Oberteuffer1970,Sliwko1994},
and can grow in different crystalline phases. X-ray diffraction and resistivity measurements indicate that our Mn films are polycrystalline and disordered \cite{SuppPR}. The unusual temperature dependence of the Mn resistance in Fig. 4(b) is another indication of the glass-like metallic behavior of Mn thin films, consistently with previous work \cite{Grassie1979,Boakye1999,Mooij1973}. 
Because the transport of orbital momentum depends strongly on the coherence of the orbital part of the electronic wavefunction, electron scattering at grain boundaries and crystal defects can be major sources of the orbital quenching and relaxation. In agreement with this hypothesis, the thickness dependence of the Mn resistivity reported in Ref.~\onlinecite{SuppPR} suggests an electron scattering length shorter than 5~nm. Therefore, the complex and disordered crystalline structure of Mn may be the limiting factor of the orbital generation and diffusion.
The anomalous resistive behavior associated with the structural complexity of Mn is likely also the reason for the temperature dependence of the HMR reported in Fig.~4(b). Differently from the HMR of Pt \cite{Velez2016}, the HMR of Mn decreases at low temperature. Such a decrease may be ascribed, at least in part, to the reduced low temperature conductivity typical of disordered metal systems \cite{Mooij1973,Kaveh1982}. In future studies, relating the HMR, its temperature dependence, and the orbital parameters to the crystalline structure and disorder of metal films may help to understand the mechanisms favoring orbital transport. In this respect, we note that $\sigma_L$ estimated from magneto-optical measurements in a single nonmagnetic Ti layer is also two orders of magnitude smaller than predicted by theory \cite{Choi2021a}. This discrepancy and the scaling of the orbital Hall angle with the longitudinal conductivity \cite{Hayashi2023} suggest that extrinsic mechanisms could possibly be responsible for the orbital generation. So far, however, there are no predictions of an extrinsic orbital Hall effect nor estimates of the influence of the electronic scattering on the orbital transport.

\begin{figure}[t]
\includegraphics[scale=1]{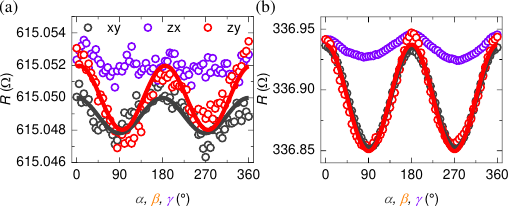}
\caption{(a) Longitudinal resistance of BiYIG/Mn(10) measured by rotating a constant magnetic field of 1 T in the $xy$, $zx$, and $zy$ planes. The red and grey lines are fits of $\cos^2(\alpha, \beta)$ to the data. (b) Same as (a) in BiYIG/Pt(5).}
\label{fig:fig5}
\end{figure}

In comparison to Mn, we find that Pt has a slightly higher diffusion coefficient ($D \approx 4.5\cdot 10^{-6}$ m$^2$/Vs), and a shorter relaxation time ($\tau <$ 1 ps). These values are in agreement with previous results \cite{Velez2016}, and explain the parabolic dependence of the magnetoresistance in Fig. 3(c). The combination of a large diffusion coefficient and a small relaxation time implies that stronger magnetic fields are necessary to cause the dephasing of the angular momentum and, hence, modulate the sample resistance.

In addition to these considerations, further evidence of the orbital nature of the magnetoresistance in Mn is provided by SMR measurements in bilayers of BiYIG/Mn and BiYIG/Pt \cite{SuppPR}. Because $\theta$ of Mn and Pt differ by a factor $\approx$ 2-3, the SMR in BiYIG/Mn should be 5-10 times smaller than in BiYIG/Pt if only spin accumulation occurred at the interface. This prediction, however, is at odds with our observation [Fig. 5]. In BiYIG/Pt(5), the normalized SMR is of the order of 2.6$\cdot 10^{-4}$, which is similar to earlier reports \cite{Althammer2013}. In contrast, the SMR in BiYIG/Mn(10) is much weaker. Fitting a $\cos^2\beta$ function to the $zy$ angle scan yields an SMR of about 6.4$\cdot 10^{-6}$, which is 40 times smaller than in BiYIG/Pt(5). In comparison, the HMR in BiYIG/Mn(10) ($\approx 4.5\cdot 10^{-5}$) and BiYIG/Pt(5) ($\approx 1.5\cdot 10^{-4}$) are similar to those found in Mn and Pt single layers grown on SiO$_2$ \cite{SuppPR}. In principle, the different SMR amplitude in BiYIG/Mn and BiYIG/Pt could be assigned to a significant variation of the spin-mixing conductance between the two samples. However, previous works have found similar mixing conductance at the YIG/Mn and YIG/Pt interfaces \cite{Du2014,Qu2018,Du2015}. Therefore, the similar and large HMR but very different SMR in Mn and Pt indicate that different actors are at play in the two elements. In Pt, the spin accumulation determined by the strong spin-orbit coupling is responsible for both the HMR and SMR because the spin angular momentum couples to both the external magnetic field (HMR) and the interfacial exchange field (SMR). In contrast, orbital moments precess and dephase in the magnetic field, but do not interact directly with the magnetization \cite{Go2021}. This means that a pure orbital accumulation can generate a finite HMR, but cannot produce any SMR. Therefore, we ascribe the magnetoresistance of Mn to an orbital-driven effect.
In this scenario, a small but non-zero spin orbit coupling may generate the tiny SMR detected in BiYIG/Mn(10).

In conclusion, we have observed a strong Hanle effect in Mn thin films. The combination of an unexpected large HMR with an almost zero SMR in a 3$d$ element indicates that this magnetoresistance originates from orbital moments rather than spin moments. Similar orbital magnetoresistance effects may exist in other transition metals. Additionally, we find an orbital Hall angle of about 0.016, an  orbital relaxation time of about 2 ps and an orbital diffusion length of the order of 2 nm. The small orbital Hall conductivity and short diffusion length are at odds with the expected strength and lengthscale of orbital transport and associated to the disordered structure of Mn thin films. 
The Hanle effect therefore provides a new means to explore the orbital physics in nonmagnetic elements. 

\begin{acknowledgments}
We thank Dongwook Go, Mingu Kang, Paul Nöel, and Richard Schlitz for providing useful comments. This research was supported by the Swiss National Science Foundation (Grant No. 200020-200465). Hanchen Wang acknowledges the support of the China Scholarship Council (CSC, Grant No. 202206020091). William Legrand acknowledges the support of the ETH Zurich Postdoctoral Fellowship Program (21-1 FEL-48).
\end{acknowledgments}

\section{Bibliography}

\appendix

\bibliography{library.bib}

\end{document}